\documentclass[preprint,showpacs,preprintnumbers,amsmath,amssymb,nofootinbib]{revtex4}
\usepackage{bbm}
\usepackage{amsfonts}
\usepackage{booktabs}
\usepackage{mathrsfs}
\usepackage{epsfig}
\usepackage{graphicx}% Include figure files
\usepackage{dcolumn}% Align table columns on decimal point
\usepackage{bm}% bold math
\usepackage{amsmath}
\usepackage{slashed}
\usepackage{subfigure}

\let\jnfont=\rm
\def\NPB#1,{{\jnfont Nucl.\ Phys.\ B }{\bf #1},}
\def\PLB#1,{{\jnfont Phys.\ Lett.\ B }{\bf #1},}
\def\EPJC#1,{{\jnfont Eur.\ Phys.\ Jour.\ C }{\bf #1},}
\def\PRD#1,{{\jnfont Phys.\ Rev.\ D }{\bf #1},}
\def\PRL#1,{{\jnfont Phys.\ Rev.\ Lett.\ }{\bf #1},}
\def\MPLA#1,{{\jnfont Mod.\ Phys.\ Lett.\ A }{\bf #1},}
\def\JPG#1,{{\jnfont J.\ Phys.\ G}{\bf #1},}
\def\CTP#1,{{\jnfont Commun.\ Theor.\ Phys.\ }{\bf #1},}
\def\ZPC#1,{{\jnfont Z.\ Phys.\ C }{\bf #1},}
\def\JHEP#1,{{\jnfont JHEP \ }{\bf #1},}
\def\Rv{\not{\hbox{\kern-1pt $R$}}}
\def\p{\not{\hbox{\kern-3pt $p$}}}

\begin{document}

\title{Probing Light Higgsinos in Natural SUSY from Monojet Signals at the LHC}% Force line breaks with \\
%%\thanks{A footnote to the article title}%

\author{Chengcheng Han$^{1,2}$, Archil Kobakhidze$^3$, Ning Liu$^{1,3}$, Aldo Saavedra$^3$}
\author{Lei Wu$^3$}
\author{Jin Min Yang$^2$}
\affiliation{$^1$ Physics Department, Henan Normal University, Xinxiang 453007, China\\
$^2$ State Key Laboratory of Theoretical Physics, Institute of Theoretical Physics,
 Academia Sinica, Beijing 100190, China\\
$^3$ ARC Centre of Excellence for Particle Physics at the Terascale, School of Physics,
The University of Sydney, NSW 2006, Australia
}%

\date{\today}% It is always \today, today,

\begin{abstract}

We investigate a strategy to search for light, nearly degenerate higgsinos within the natural MSSM at the LHC. We demonstrate that the higgsino mass range $\mu$ in $100-150$ GeV, which is preferred by the naturalness, can be probed at $2\sigma$ significance through the monojet search at 14 TeV HL-LHC with 3000 fb$^{-1}$ luminosity. The proposed method can also probe certain region in the parameter space for the lightest neutralino with a high higgsino purity, that cannot be reached by planned direct detection experiments at XENON-1T(2017).

\end{abstract}

\maketitle

%\tableofcontents

\section{\label{sec:level1}Introduction}
One of the key theoretical motivations for low-energy supersymmetry (SUSY) is that it provides a framework for a light Higgs boson  without invoking unnatural fine-tuning of theory parameters. However, recent discovery of a Standard Model (SM) Higgs-like particle with the
mass around 125 GeV \cite{atlas, cms}, in conjunction with non-observation of supersymmetric particles, have largely excluded the most studied parameter range within the minimal supersymmetric Standard Model (MSSM), for which the naturalness criterion is satisfied. If the observed resonance is to be identified with the lightest CP-even Higgs boson of MSSM, heavy multi-TeV stops and/or large Higgs-stop trilinear soft-breaking  coupling are required to achieve sufficient enhancement of the predicted Higgs mass \cite{mhiggs, mssm-stop}. Furthermore, null results on gluino searches at the LHC so far have pushed the lower limit on gluino mass above the TeV scale \cite{gluino}. All these significantly jeopardize the naturalness of MSSM with a standard sparticle spectrum. Therefore, it is imperative to consider the possibly hidden parameters space where the theory maintains naturalness, and look for other strategies for verifying such natural SUSY models at the LHC \cite{naturalness}. In this work, we investigate the possibility of monojet signals induced by light higgsinos at 14 TeV high-luminosity LHC(HL-LHC) as a probe of natural SUSY.

The justification for light, nearly degenerate higgsinos within the natural MSSM comes from the following consideration.
In the MSSM, the minimization of the tree-level Higgs potential leads to the relation \cite{mz}:
\begin{eqnarray}
\frac{M^2_{Z}}{2}=-\mu^{2}+\frac{m^2_{H_d}-m^2_{H_u}\tan^{2}\beta}{\tan^{2}\beta-1}\simeq-\mu^{2}-m^2_{H_u},
\label{minimization}
\end{eqnarray}
where $m^2_{H_d}$ and $m^2_{H_u}$ represent the weak scale soft SUSY breaking masses of
the Higgs fields, and $\mu$ is the higgsino mass parameter. A moderate/large
$\tan\beta \gtrsim 10$ is assumed in the last approximate
equation. In order to avoid large fine-tuning in Eq.(\ref{minimization}), $\mu$ and $m_{H_u}$ must be of the order of $\sim 100-200$ GeV, which implies light higgsinos. At the same time, the electroweak gaugino mass parameters $M_{1,2}$ are preferred to be of the similar order as the heavy gluino mass parameter $M_{3}$ and large Higgs-stop trilinear coupling $A_t$ is needed \cite{non-universal-gaugino}. Hence, generically we have $\mu \ll M_{1,2}$ and the mass splittings between the lightest chargino and the lightest two neutralinos at leading order are determined by \cite{giudice}
\begin{eqnarray}
 m_{\tilde{\chi}^\pm_1} - m_{\tilde{\chi}^0_1} &=&
  \frac{M_W^2}{2 M_2} \left( 1 - \sin 2\beta - \frac{2 \mu}{M_2} \right) \nonumber  \\
&& + \frac{M_W^2}{2 M_1} \tan^2\theta_W (1 + \sin 2\beta) , \\
m_{\tilde{\chi}^0_2} - m_{\tilde{\chi}^0_1} &=&
  \frac{M_W^2}{2 M_2} \left( 1 - \sin 2\beta + \frac{2 \mu}{M_2} \right) \nonumber \\
&& + \frac{M_W^2}{2 M_1} \tan^2\theta_W (1 - \sin 2\beta) \; .
\end{eqnarray}
This in turn implies that light electroweak gauginos in the natural MSSM are nearly degenerate higgsino-like states with a mass differences of about $3-10$ GeV (for $M_1=M_2\sim0.5-2$ TeV). Therefore, a direct search for light higgsinos may serve as a sensitive probe of the natural MSSM.

For such light higgsinos the electroweak production rates for $Z \to \tilde\chi^0_1\tilde\chi^0_1$
and  $Z \to \tilde\chi^0_2\tilde\chi^0_2$ are suppressed, while the production rates for
$Z/\gamma^* \to \tilde\chi^+_1\tilde\chi^-_1$, $Z\to\tilde\chi^0_1\tilde\chi^0_2$,
$W^{\pm} \to \tilde\chi^{\pm}_1\tilde\chi^0_1$ and $W^{\pm} \to \tilde\chi^{\pm}_1\tilde\chi^0_2$
are expected to be reasonably large, reaching pb-level at the LHC. However, since the light
higgsinos are nearly degenerate, the products of their subsequent decays,
$\tilde{\chi}^\pm_1 \to W^{\pm*} \tilde\chi^{0}_{1}$ and $\tilde{\chi}^0_2 \to Z^{*}\tilde{\chi}^0_1$,
will carry small energies and, hence, the currently adopted search strategy for electroweak
gauginos through their direct pair production is not applicable to this case\cite{gunion,liantao}. Recently, a new search
channel based on the wino pair production with a same-sign diboson plus missing transverse
energy ($\slashed E_T$) final state has been proposed for the 14 TeV LHC in \cite{wino-pair}.
Also, it has been pointed out that for $(m_{\tilde\chi^\pm_1} - m_{\tilde\chi^0_1})\lesssim 1$ GeV the wino may have a long life-time
and such long-lived charged particle is already excluded by the LHC data \cite{long-lived}.

%%%fig1.eps%%%%%%%%%%%%%%%%%%%%
\begin{figure}[h]
\centering
\includegraphics[width=3in,height=1.2in]{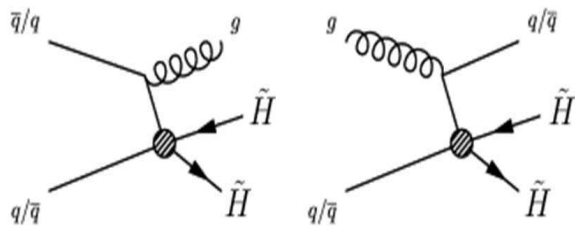}
\vspace{-0.3cm}
\caption{Feynman diagrams depicting monojet production in the natural MSSM at the LHC.}
\label{spectrum}
\end{figure}
%%%%%%%%%%%%%%%%%%%%%%%%%%%%%%%%

\section{\label{sec:level1}Calculations and discussions}

We study the detection of the light higgsinos via monojet searches at the LHC in the following processes (see Fig. 1 for the corresponding Feynman diagrams ):
\begin{eqnarray}
pp \to \tilde{\chi}^{\pm}_{1}\tilde{\chi}^{\mp}_{1}j,~~\tilde{\chi}^{0}_{1}\tilde{\chi}^{0}_{2}j,~~\tilde{\chi}^{\pm}_{1}\tilde{\chi}^{0}_{1,2}j.\label{signal}
\end{eqnarray}
In these processes a hard jet radiated from initial partons recoils against the invisible missing transverse energy from soft decay productes and this can be used as a handle to tag the higgsino pair production. Because of the small mass splitting ($\Delta m \sim 3-10$ GeV) between $\tilde\chi^{\pm}_{1},\tilde\chi^{0}_{2}$ and $\tilde\chi^{0}_{1}$, all three channels ($j\tilde{\chi}^{+}_{1}\tilde{\chi}^{-}_{1}$,$j\tilde{\chi}^{0}_{1}\tilde{\chi}^{0}_{2}$ and $j\tilde{\chi}^{\pm}_{1}\tilde{\chi}^{0}_{1,2}$) share the same topology in the detector. As a result, the monojet production rates within the natural MSSM are greatly enhanced. In addition, when $\mu \ll M_{1,2}$, these processes are largely insensitive to other SUSY parameters but higgsino mass $\mu$. Therefore, we do not consider the production of stops and gluino in this paper, which contribute to the fine-tuning in more complicated and model-dependent way \cite{non-universal-gaugino}. The current constraints on the mass limits of stop and gluino in natural SUSY have been discussed in \cite{direct-stop,stop-constraint}. The sleptons and first two generation squarks are irrelevant for our analysis and we assume them to be heavy.

Since the monojets have a distinctive topology of events with a singly high $p_T$ hadronic jets and large missing $\slashed E_T$, their relevance to the search for the pair production of weakly-interacting particles have been exploited at the LHC \cite{lhc}. The SM backgrounds to the above monojet signature are dominated by the following four processes: (i) $pp \to Z(\to \nu\bar{\nu})+j$, which is the main irreducible background with the same topology as our signals; (ii) $pp \to W(\to \ell \nu)+j$, this process fakes the signal only when the charged lepton is outside the acceptance of the detector or close to the jet; (iii) $pp \to W(\to \tau \nu)+j$, this process may fake the signal since a secondary jet from hadronic tau decays tend to localize on the side of $\slashed E_T$; (iv) $pp \to t\bar{t}$, this process may resemble the signal, but also contains extra jets and leptons. This allows to highly suppress $t\bar{t}$ background by applying a b-jet, lepton and light jet veto.

For the QCD background, the misreconstruction of the energy of a jet in the calorimeters
can cause an ordinary di-jet event with large missing energy to mimic the signal.
An estimation of the QCD background based on the full detector simulation can be found
in  \cite{multijet}. By fitting the jet energy response function (JERF) using the method in \cite{jerf},
the authors of \cite{allanach} found that the multijet background in the supersymmetric monojets analysis at 14 TeV
LHC can be reduced to a negligible level by requiring a large $\slashed E_T$ cut,
such as $\slashed E_T > 200$ GeV . Since other dominant backgrounds have $\slashed E_T > 200$ GeV,
we set $\slashed E_T > 500$ GeV as in \cite{drees}, where the cuts for the monojet events are optimized for 14 TeV LHC, thus we can safely neglect the QCD background in our calculation (The pile-up effects at 14 TeV HL-LHC have not been considered in the work, due to lack of the exact detector configurations.).
The diboson backgrounds and single top background are not considered in our calculations
due to their small cross sections compared to other backgrounds.

%%%fig2.eps%%%%%%%%%%%%%%%%%%%%
\begin{figure}[h]
\centering
\includegraphics[width=3in,height=2.5in]{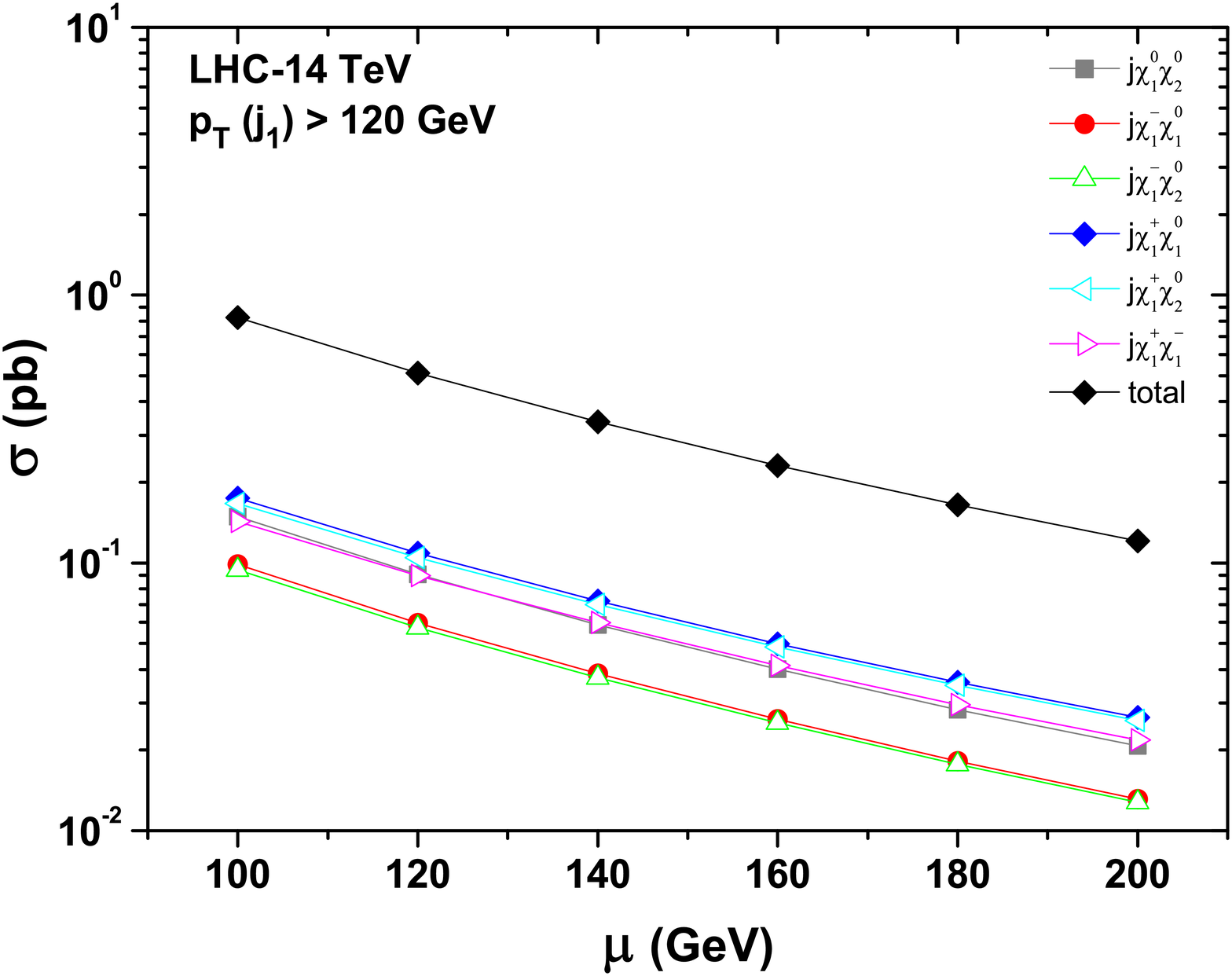}
\vspace{-0.3cm}
\caption{The parton-level cross section of monojet signals at 14 TeV LHC.}
\label{xsection}
\end{figure}
%%%%%%%%%%%%%%%%%%%%%%%%%%%%%%%

In the calculations we assume $M_1=M_2=1$ TeV and use the \textsf{Suspect} \cite{suspect} and \textsf{SUSY-HIT} \cite{susyhit} to calculate masses, couplings and branching ratios of the relevant sparticles. The parton level signal and background events are generated with \textsf{MadGraph5} \cite{mad5}. We perform parton shower and fast detector simulations with \textsf{PYTHIA} \cite{pythia} and \textsf{Delphes} \cite{delphes}. We cluster jets using the anti-$k_t$ algorithm with a cone radius $\Delta R=0.7$ \cite{anti-kt}. In order to obtain reasonable statistics, a generator level event filter was applied which imposed a parton-level cut of $p_T > 120$ GeV on the first leading jet for signals and $W/Z+j$ backgrounds.
It should be noted that the jet veto cuts can significantly affect the QCD corrections to the backgrounds \cite{denner}.
To include the QCD effects, we generate parton-level events of $Z/W+j$ with up to two jets that are matched to the parton
shower using the MLM-scheme with merging scale $Q=60$ GeV \cite{mlm}. Due to the $t\bar t$ events containing a large number
of jets, we need not generate the events with the extra hard partons, which will be strongly rejected by the jets veto
\cite{drees}. Although the additional jet may come from the decays of $\tilde{\chi}^{\pm}_{1}$ or $\tilde{\chi}^{0}_{2}$,
they are too soft to pass our strict $p_T$ cut on the leading jet adopted in the following analysis. So there is no need
to generate the higgsinos pairs without additional parton in the final state. Besides, our signal simulation is exclusively
based on Eq.(\ref{signal}) so that double counting will not arise in our calculation.

%%%fig2.eps%%%%%%%%%%%%%%%%%%%%
\begin{figure}[h]
\centering
\includegraphics[width=3in,height=2.5in]{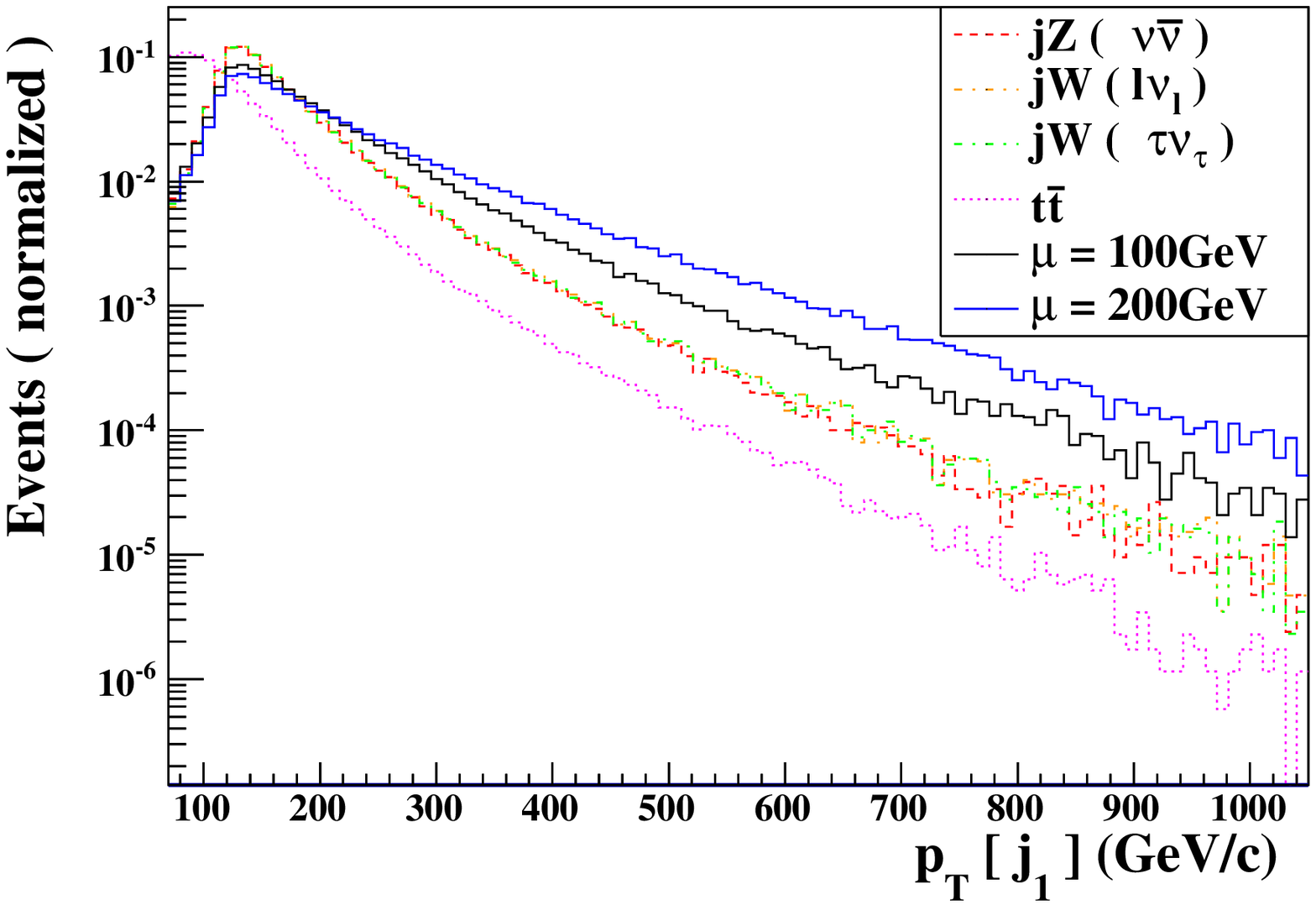}
\includegraphics[width=3in,height=2.5in]{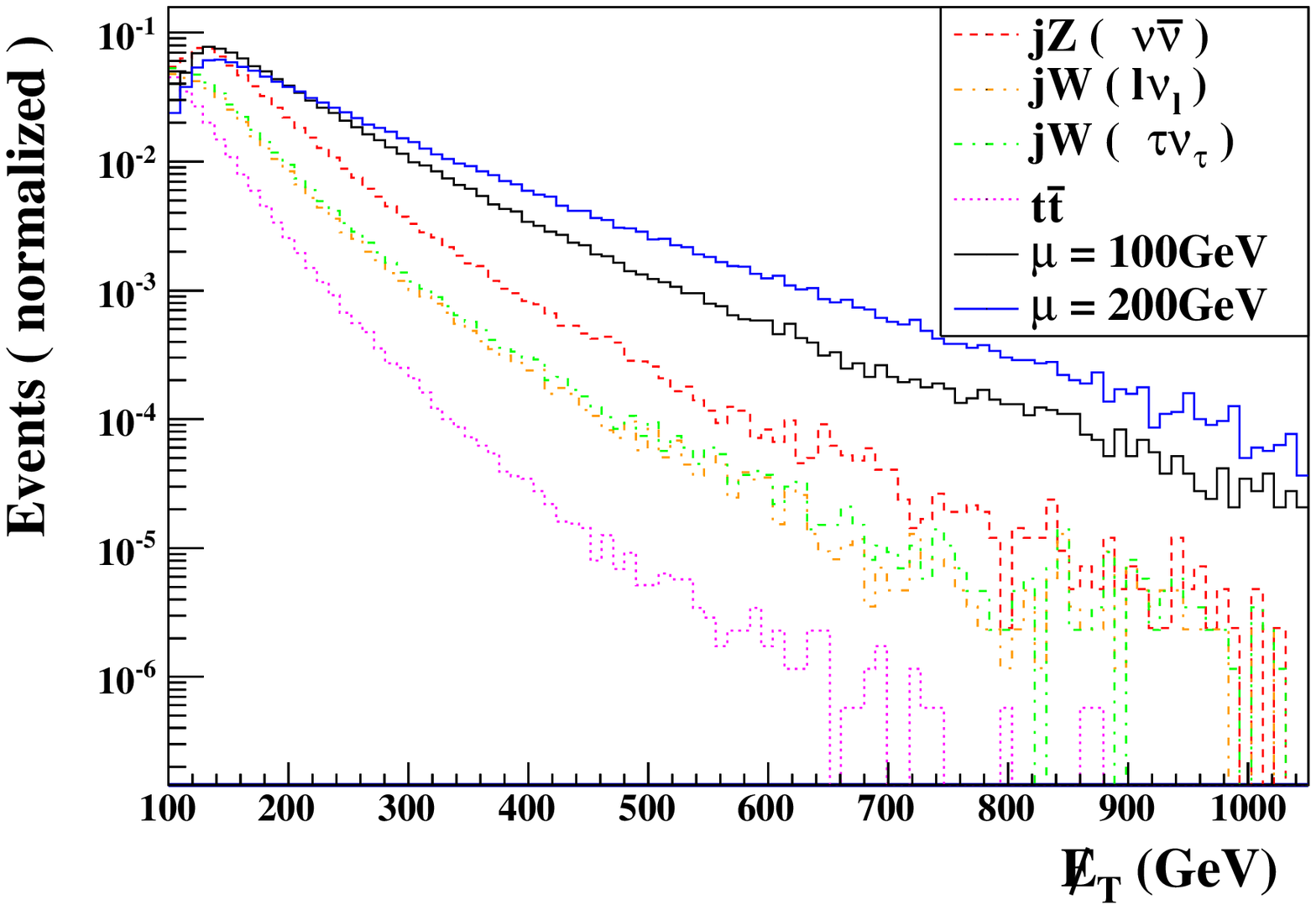}
\caption{The normalized distributions of the reconstructed leading jet $p_T(j_1)$ and $\slashed E_T$ of the monojet signals and backgrounds at 14 TeV LHC.}
\label{pt}
\end{figure}
%%%%%%%%%%%%%%%%%%%%%%%%%%%%%%%%

In Fig.\ref{xsection}, we display the cross section of $pp \to \tilde{\chi}^{\pm}_{1}\tilde{\chi}^{\mp}_{1}j,~\tilde{\chi}^{0}_{1}\tilde{\chi}^{0}_{2}j,~\tilde{\chi}^{\pm}_{1}\tilde{\chi}^{0}_{1,2}j$ as a function of higgsino mass $\mu$ after requiring the parton-level cut $p_T(j_1) > 120$ GeV at 14 TeV LHC. Since $ug$ initial states have large parton distribution function, the largest contribution to the cross section of our signals comes from $\tilde{\chi}^{+}_{1}\tilde{\chi}^{0}_{1}j$. The degenerate spectrum of $\tilde{\chi}^{\pm}_{1}$ and $\tilde{\chi}^{0}_{1,2}$ implies that signals with the same initial states have approximately same cross sections. Therefore, the total production rate is amplified and can reach nearly pb-level.

In Fig.\ref{pt} we show the normalized distributions of a reconstructed leading jet $p_T(j_1)$ and $\slashed E_T$ of the signals and backgrounds. From the upper panel one can see that for $p_T(j_1)>200$ GeV the signals have harder $p_T(j_1)$ spectrum  than the backgrounds. The greater value of $\mu$ corresponds to an increase in the average $p_T$ of the jet. The difference in peaks of the signals ($\sim 120$ GeV) and $t\bar{t}$ background ($\sim m_t/2$) is caused by the parton-level cut $p_T(j_1)>120$ GeV. From the lower panel one observes that the signals have the larger $\slashed E_T$ than the backgrounds. Thus, a hard cut on $\slashed E_T$ will be effective to reduce the backgrounds.

According to the above analysis, events are selected to satisfy the following criteria of monojet searches \cite{lhc}, and the cuts for $\slashed E_T$ and $p_T(j_1)$ are optimized for 14 TeV LHC \cite{drees}: $(i)$ We require large missing transverse energy $\slashed E_T > 500$ GeV; $(ii)$ The leading jet is required to have $p_T(j_1) > 500$ GeV and $|\eta_{j_1}| < 2$; events with more than two jets with $p_T$ above 30 GeV in the region $|\eta| < 4.5$ are rejected; $(iii)$ We veto the second leading jet with $p_T(j_2) > 100$ GeV and $|\eta_{j_2}| < 2$; $(iv)$ A veto on events with an identified lepton ($\ell=e,\mu,\tau$) or $b$-jet is imposed to reduce the background of $W+j$ and $t\bar{t}$. We use the $b$-jet tagging efficiency parametrisation given in \cite{cms-b} and include a misidentification 10\% and 1\% for $c$-jets and light jets respectively. We also assume the $\tau$ tagging efficiency is 40\% and include the mis-tags of QCD jets by using \textsf{Delphes}.

\begin{table*}[t!]
\begin{center}
\begin{ruledtabular}\scriptsize
\begin{tabular}{c||c|c|c|c|c|c}
cut & $Z(\nu\bar \nu)+j$ &
$W(\ell \nu_\ell)+j$ &
$W(\tau\nu_\tau)+j$ & $t\bar t$  &Signal ($\mu=100$ GeV) & Signal ($\mu=200$ GeV) \\
\hline
$p_T(j_1)>500 \rm GeV$ & 69322 & 241740 & 119078 & 210943 & 1242 & 415\\\hline
$\slashed E_T >500 \rm GeV$ & 26304 & 28209 & 16513 &  2786 & 950 & 335\\ \hline
veto on $p_T(j_2)>100$, $p_T(j_3)>30$ & 16988 & 12194 &  7577 & 306 & 602 & 223\\\hline
veto on $e, \, \mu, \, \tau$ & 16557 & 3963 & 3088 & 102 & 597 & 220\\\hline
veto on $b-$jets & 16303 & 3867 & 3046 & 56 & 576 & 214
\end{tabular}
\end{ruledtabular}
\caption{Cut flow of the signal events for $\mu=100,200$ GeV at 14 TeV LHC with
$\mathcal{L} = 100\,\rm{fb}^{-1}$. The cross section of $t\bar{t}$ is normalized
to the approximately next-to-next-to-leading order value $\sigma_{t\bar{t}}=920$ pb \cite{ttbar}.
\label{tab:cut_flow}}
\end{center}
\normalsize
\end{table*}

In Table \ref{tab:cut_flow}, the resulting cut-flow for signal and background events is presented, for a centre-of-mass energy of 14 TeV and an integrated luminosity of 100 fb$^{-1}$. After the cuts $P_T(j_1)>500$ GeV and $\slashed E_T >500$ GeV, the $Z+j$ and $W+j$ backgrounds are reduced by $\mathcal{O}(10^{-4})$, while the signals only by $\mathcal{O}(10^{-2})$. The lepton and light jet veto will suppress $Wj$ backgrounds by extra two orders.
For $t\bar{t}$ background, we have not included the hadronic channels due to its large jet multiplicity and small $\slashed E_T$. We impose the third jet veto as the requirement of the ATLAS collaboration \cite{lhc}, which is not used in the paper \cite{drees}. We also checked that our results are consistent with those obtained in Ref. \cite{nojiri} by setting the same values of cuts and collider energy. The $Z(\nu\bar{\nu})$ process is still the dominant background after all cuts.

%%%fig3.eps%%%%%%%%%%%%%%%%%%%%
\begin{figure}[h]
\centering
\includegraphics[width=3in,height=2.5in]{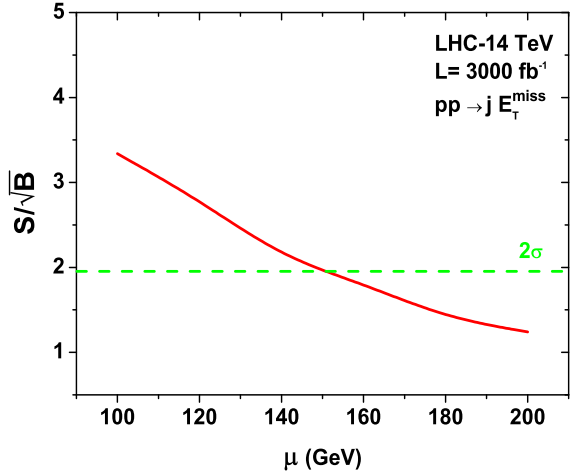}
\vspace{-0.3cm}
\caption{The dependence of significance on the higgsino mass $\mu$ at 14 TeV HL-LHC with $\mathcal{L} = 3000$ fb$^{-1}$.}
\label{ss}
\end{figure}
%%%%%%%%%%%%%%%%%%%%%%%%%%%%%%%%

In Fig.\ref{ss} we display the dependence of the signal significance $S/\sqrt{B}$ on the higgsino mass $\mu$ at 14 TeV HL-LHC for various luminosities, $\mathcal{L} = 3000$ fb$^{-1}$. The overall background $B$ including the systematic errors is calculated through the formula $B = \sum_i B_{i}+\sum_i (0.01 B_{i})^2, (i= Z+j, t\bar t\rm{, } \,W(\rightarrow \ell\nu_\ell)+j \rm{, } \,W(\rightarrow \tau \nu_\tau)+j)$, where we assume the systematic error to be $1\%$. With an increase of $\mu$ the significance drops fast due to the reduction in the signal cross sections. At $\mathcal{L} = 3000$ fb$^{-1}$, the range $\mu \sim 100-150$ GeV, favored by the naturalness, can be probed at $2\sigma$ significance. However, it should be mentioned that, since the realistic detector performances of the HL-LHC are still not available, we can expect our analysis can be improved by optimizing signal extraction strategies and better understanding of the backgrounds uncertainties through the dedicated analysis of the experimental collaborations at HL-LHC.

%%%fig4.eps%%%%%%%%%%%%%%%%%%%%
\begin{figure}[h]
\centering
\includegraphics[width=3in,height=2.5in]{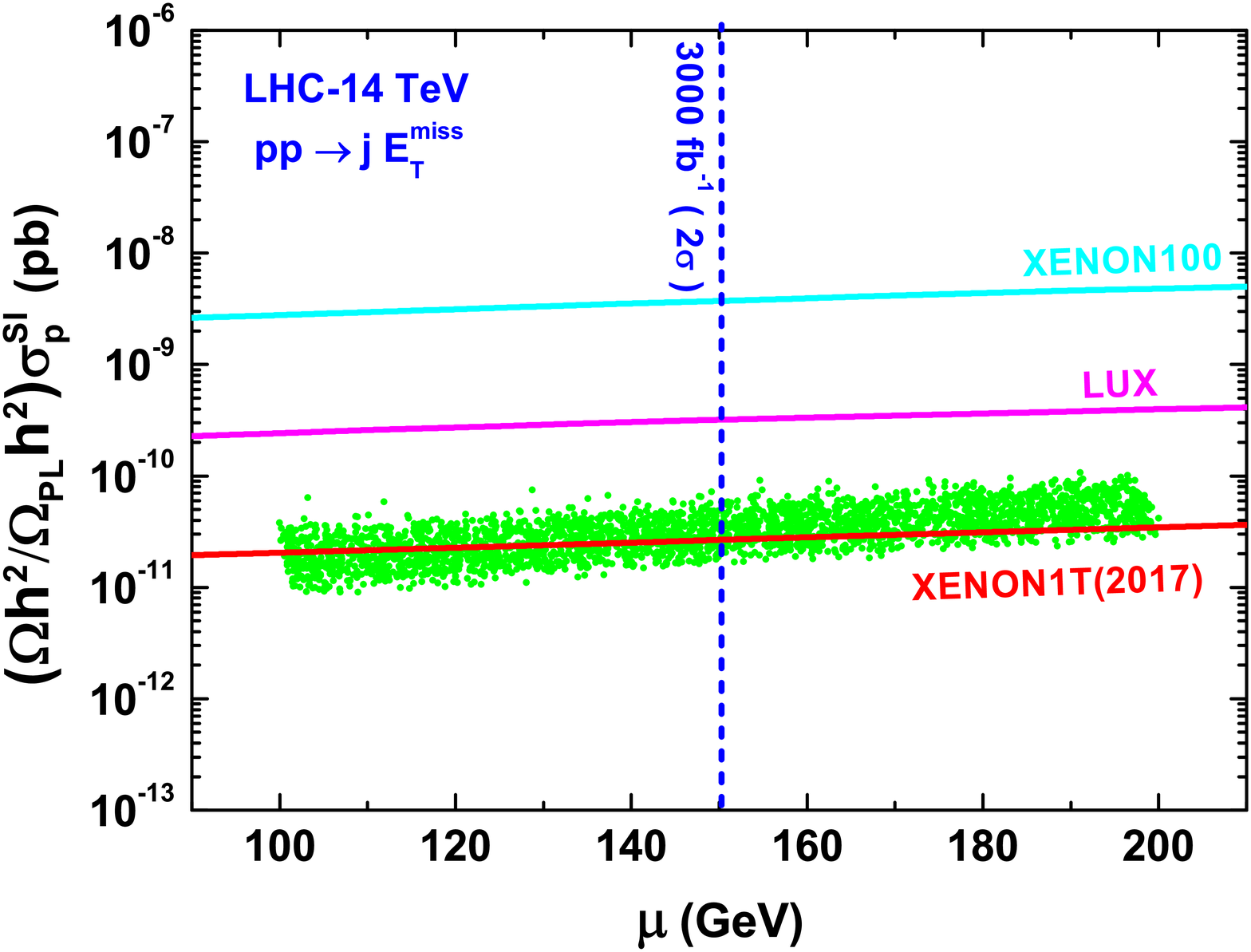}
\vspace{-0.3cm}
\caption{ Scatter plot of samples survived the constraints from (1)-(6) in the text. The horizontal lines show the 90\%~C.L. bound from XENON100 \cite{xenon100}, future sensitivities at LUX \cite{lux} and XENON1T \cite{xenon1t}, respectively. The vertical dashed line is the sensitivity of monojet signals at $2\sigma$ significance at 14 TeV LHC with $\mathcal{L} = 3000$ fb$^{-1}$.}
\label{xenon}
\end{figure}
%%%%%%%%%%%%%%%%%%%%%%%%%%%%%%%%

As a complementary searches for the light higgsinos, we also investigate the probing ability of the dark matter direct detections. We computed the dark matter observables by using the package \textsf{MicrOmega} \cite{micromega} and scan the following parameter space:
$100~{\rm GeV} \le \mu \le 200~{\rm GeV}$, $0.6~{\rm TeV} \le m_{\tilde{Q}_{L3}}$, $m_{\tilde{t}_R}=~m_{\tilde{b}_R}\le 2~{\rm TeV}$, $-3{\rm TeV}\le A_{t}=A_{b}\le3{\rm TeV}$, $1 \le \tan\beta \le 60$, $1~{\rm TeV} \le M_1,M_2 \le 2~{\rm TeV}$. Other irrelevant mass parameters are taken as 2 TeV. The above parameters are further constrained by: (1) Measurements of $B\rightarrow X_s\gamma$ and $B_s\rightarrow \mu^+\mu^-$ processes at $2\sigma$ level \cite{superiso}; (2) Higgs mass in the range 123-127 GeV \cite{feynhiggs}; (3) LHC searches for $H/A \to \tau^+\tau^-$ \cite{cms-htata}; (4) Direct search results of stop/sbottom pair productions at the LHC \cite{stop-constraint}; (5) LEP data \cite{higgsbounds} and (6) Electroweak precision measurements \cite{rb}.

We note that, in the natural MSSM, the thermal relic density of the light higgsino-like
neutralino dark matter is typically low due to the large annihilation rate in the early universe.
This makes the standard thermally produced WIMP dark matter inadequate
in the natural MSSM.
In order to provide the required relic density,
several alternative ways have been proposed \cite{non-sm-dm1,non-sm-dm2,non-sm-dm3},
such as choosing the axion-higgsino admixture as the dark matter \cite{axion}.
In this case, the spin-independent neutralino-proton scattering
cross section $\sigma^{SI}_p$ must be re-scaled by a factor
$\Omega_{\tilde{\chi}^0_1}h^2/\Omega_{PL}h^2$ \cite{axion}, where $\Omega_{PL}h^2$ is the relic
density measured by Planck satellite \cite{planck}.
However, it should be mentioned that, if the naturalness requirement is relaxed,
the heavy higgsino-like neutralino with a mass about 1 TeV
can solely produce the correct relic density in the MSSM \cite{roy}.
Of course, all these analyses are performed by assuming a standard $\Lambda$CDM
model.

The results for the spin-independent higgsino-proton scattering cross section
are shown in Fig.\ref{xenon}  and compared with the current limits from XENON-100, LUX  \cite{xenon100, lux}
and future reach projections of XENON-1T \cite{xenon1t}. We also present the $2\sigma$ probing
sensitivity of the higgsino mass $\mu$ by our proposed monojet strategy at the LHC
with $L = 3000$ fb$^{-1}$. From Fig. \ref{xenon} we can see that even with the scale factor
$\Omega_{\tilde{\chi}^0_1}h^2/\Omega_{PL}h^2$, most of the samples can be probed by the
XENON-1T(2017). Only those samples corresponding to a neutralino with a high higgsino
purity can not be covered by the XENON-1T(2017). In this case, our proposed monojet
searches may be used to probe such a light higgsino-dominant neutralino with mass up
to $\sim 150$ GeV at 14 TeV LHC for $L = 3000$ fb$^{-1}$.
\vspace{0.5cm}

\section{\label{sec:level1}conclusion}

In this paper, we studied a strategy for searching light, nearly degenerate higgsinos in the natural MSSM. Our results showed that for $\mathcal{L} = 3000$ fb$^{-1}$, the higgsino mass range $\mu \sim 100-150$ GeV favored by the naturalness may be probed at $2\sigma$ significance through the monojet searches at 14 TeV LHC. Also, this method can probe certain area in the parameter space for the lightest neutralino with a high higgsino purity, that cannot be reached by planned direct detection experiments at XENON-1T(2017).

\section*{Acknowledgement}
This work was supported by the Australian Research Council, by the National Natural Science Foundation of China (NNSFC) under grants No. 11275245, 10821504, 11135003 and 11305049, by Specialized Research Fund for the Doctoral Program of Higher Education under Grant No.20134104120002 and by the Startup Foundation for Doctors of Henan Normal University under contract No. 11112.

\end{document}